\documentclass[a4paper,11pt]{article}
\usepackage{pos}
\usepackage{multirow}
\usepackage{graphicx}

\title{Improving $\pi\pi$ dispersive analyses and resonance determination with Forward Dispersion Relations}
\ShortTitle{Improving $\pi\pi$ dispersive analyses and resonance determination with FDR}

\author*[a]{P. Rab\'an}
\author[a]{J.R. Pel\'aez}
\author[a]{J. Ruiz de Elvira}

\affiliation[a]{Departamento de F\'isica Te\'orica and IPARCOS,\\
Universidad Complutense de Madrid, E-28040 Madrid, Spain}

\emailAdd{praban@ucm.es}
\emailAdd{jrpelaez@ucm.es}
\emailAdd{jacobore@ucm.es}
\abstract{We present preliminary results of an improved pion-pion scattering dispersive analysis that includes: a refined treatment of inelasticities, the introduction of G-waves, the extension of Forward Dispersion Relations as constraints up to 1.6 GeV, and data description up to roughly 1.8 GeV. Additionally, we impose Roy-like dispersion relations. As a result, we obtain three reliable solutions corresponding to three different datasets. From the Forward Dispersion Relation output, we extract resonance pole parameters in a parameterization-independent way using continued fractions.}

\FullConference{10th International Conference on Quarks and Nuclear Physics (QNP2024)\\
8-12 July, 2024\\
Barcelona, Spain\\}

\begin{document}
\maketitle

\section{Introduction}

Pions are the lightest hadrons and therefore they appear very frequently in final states in many hadronic processes. For this reason, a precise and reliable description of the $\pi\pi\to\pi\pi$ interactions is important in order to describe their rescattering in other processes. This has become specially relevant recently due to the high statistics for hadronic observables reached by several collaborations such as  ALICE, Babar, Belle or LHCb. Furthermore, lattice-QCD has brought a refreshed interest in pion-pion interaction. Another reason for improving the $\pi\pi$ scattering description lies on light meson spectroscopy, as it belongs to the non-perturbative QCD regime. Moreover, some open questions remain regarding the existence of glueballs or the nature of the lightest scalar resonances.

The experimental data for $\pi\pi\to\pi\pi$ were obtained in the 70's \cite{Hyams:1973zf,Grayer:1974cr,Hyams:1975mc,Kaminski:1996da,Cohen:1973yx,Losty:1973et,Hoogland:1977kt,Durusoy:1973aj}, as indirect measurements extracted from $\pi N\to\pi\pi N'$, making use of several approximations. Thus, the datasets usually present incompatibilities among themselves and large systematic errors. In addition, there are several solutions for the same experiment \cite{Hyams:1973zf,Grayer:1974cr,Hyams:1975mc,Kaminski:1996da}, which are incompatible from 0.9 GeV or 1.4 GeV, depending on the partial wave. Therefore, a meticulous analysis of all these datasets is necessary.

Furthermore, for many of the resonances produced in $\pi\pi$ below 2 GeV, the uncertainties of their parameters (mass, width and coupling) are dominated by models employed to describe meson-meson scattering data, usually by means of superpositions of different versions of Breit-Wigner parameterizations. Thus, it is necessary to provide pole parameters for those resonances in a model-independent way. This can be achieved by imposing dispersive constraints and the use of analytic continuation methods. Dispersive constraints, such as Forward Dispersion Relations (FDR) or Roy-like dispersion relations for the partial waves, are consequence of fundamental principles such as analyticity and causality, and they have been successfully applied to scattering data  (for $\pi\pi$ see \cite{GarciaMartin:2011cn} and references therein).  For $\pi\pi$ data the dispersive constraints of unsconstrained fits are generally not satisfied within errors, so they allow us to decide among different datasets and solutions. Given a dispersive description of scattering data, the resonance parameters can be obtained from the output of the FDR using analytic continuation methods \cite{Pelaez:2022qby}. We use continued fractions \cite{Schlessinger:1968vsk} to extract from our novel dispersive analysis the pole parameters of the lightest resonances produced in $\pi\pi\to\pi\pi$ process.

\section{Improving $\pi \pi$ dispersive amplitude analyses}

Here we report on our current work \cite{Pelaez:2024uav} to improve previous $\pi \pi \to \pi \pi$ dispersive amplitude analyses provided by some of us \cite{GarciaMartin:2011cn,Pelaez:2019eqa}, in order to obtain a better precision above 0.9 GeV, and to extract in a model-independent way the resonances produced in $\pi \pi$ below 2 GeV. Particularly, in \cite{GarciaMartin:2011cn} a set of parameterizations was provided for partial waves with angular momenta $\ell \leq 3$. They describe the data, fulfilling FDR up to 1.4 GeV, Roy \cite{Roy:1971tc} and GKPY \cite{GarciaMartin:2011cn} dispersion relations up to 1.1 GeV for S and P partial waves. This was possible by performing a Constrained Fit to Data (CFD) to simultaneously describe the data and satisfy the dispersive constraints within uncertainties. Afterwards, in \cite{Pelaez:2019eqa}, parameterizations for S0 and P partial waves were provided up to 1.8 GeV, known as Global parameterizations (they are global in the sense that they describe the datasets up to the highest energies available) for data solutions I, II and III (see \cite{Hyams:1973zf,Grayer:1974cr,Hyams:1975mc,Kaminski:1996da}). Those parameterizations were obtained as fits to the CFD parameterizations up to 1.4 GeV, and phenomenological fits to the three datasets above that energy. Pole parameters for the $f_0(500)$, $f_0(980)$, $f_0(1370)$, $f_0(1500)$, $f_2(1270)$ and $\rho(770)$ were obtained from those dispersive analysis in \cite{GarciaMartin:2011jx,Pelaez:2022qby}. 

\begin{figure*}[h]
\centering
\includegraphics[width=0.33\textwidth]{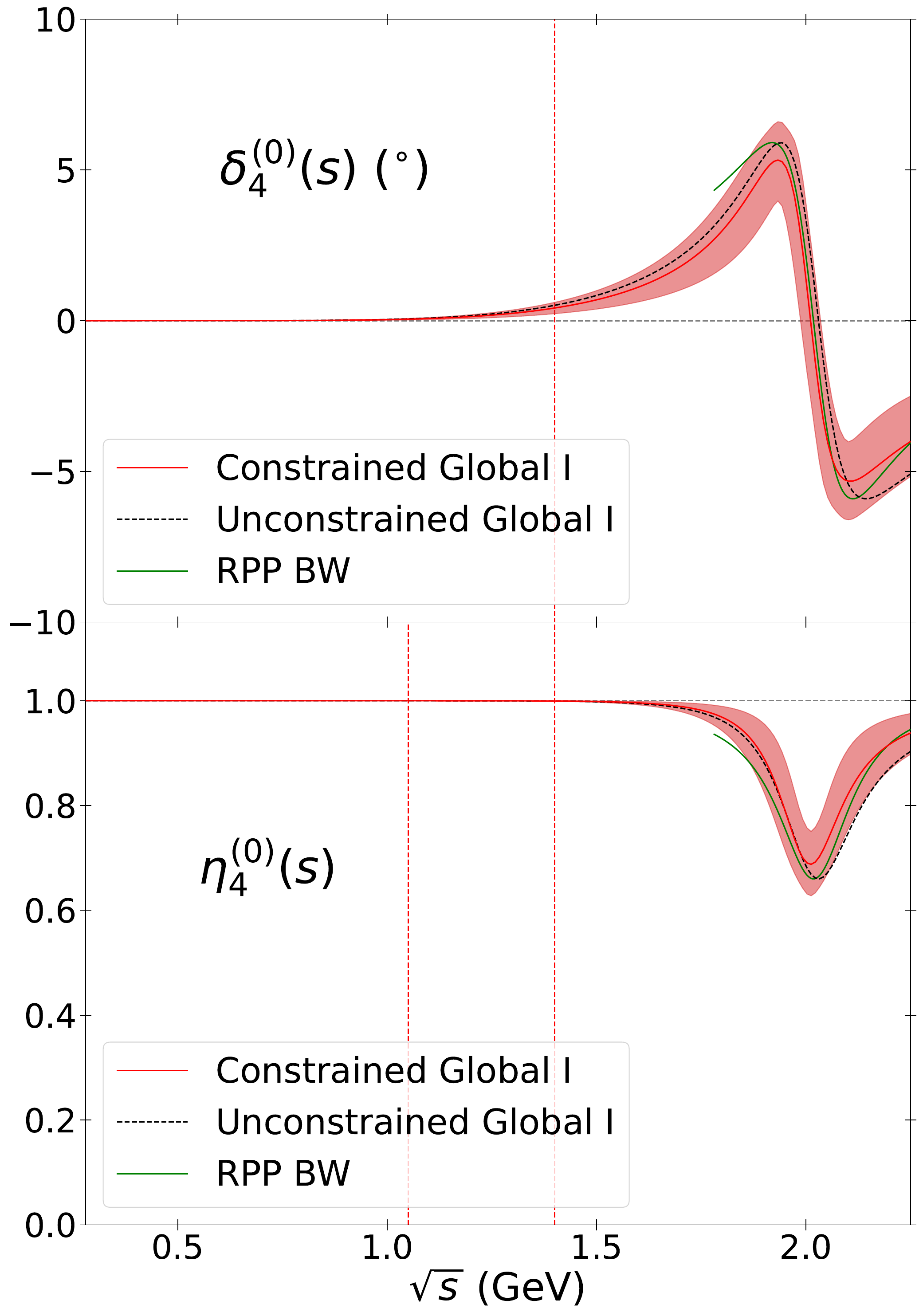}
\includegraphics[width=0.33\textwidth]{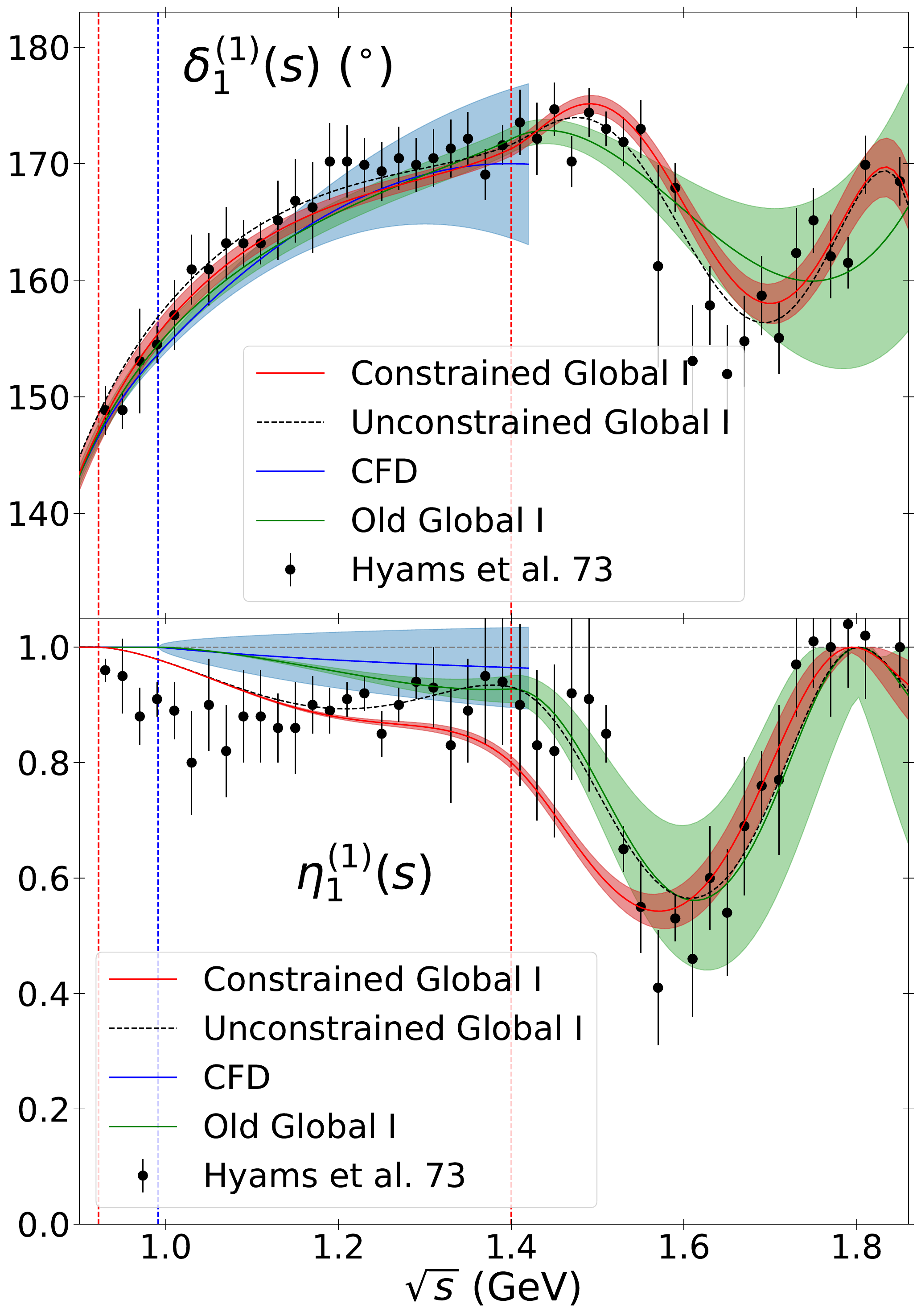}
\includegraphics[width=0.32\textwidth]{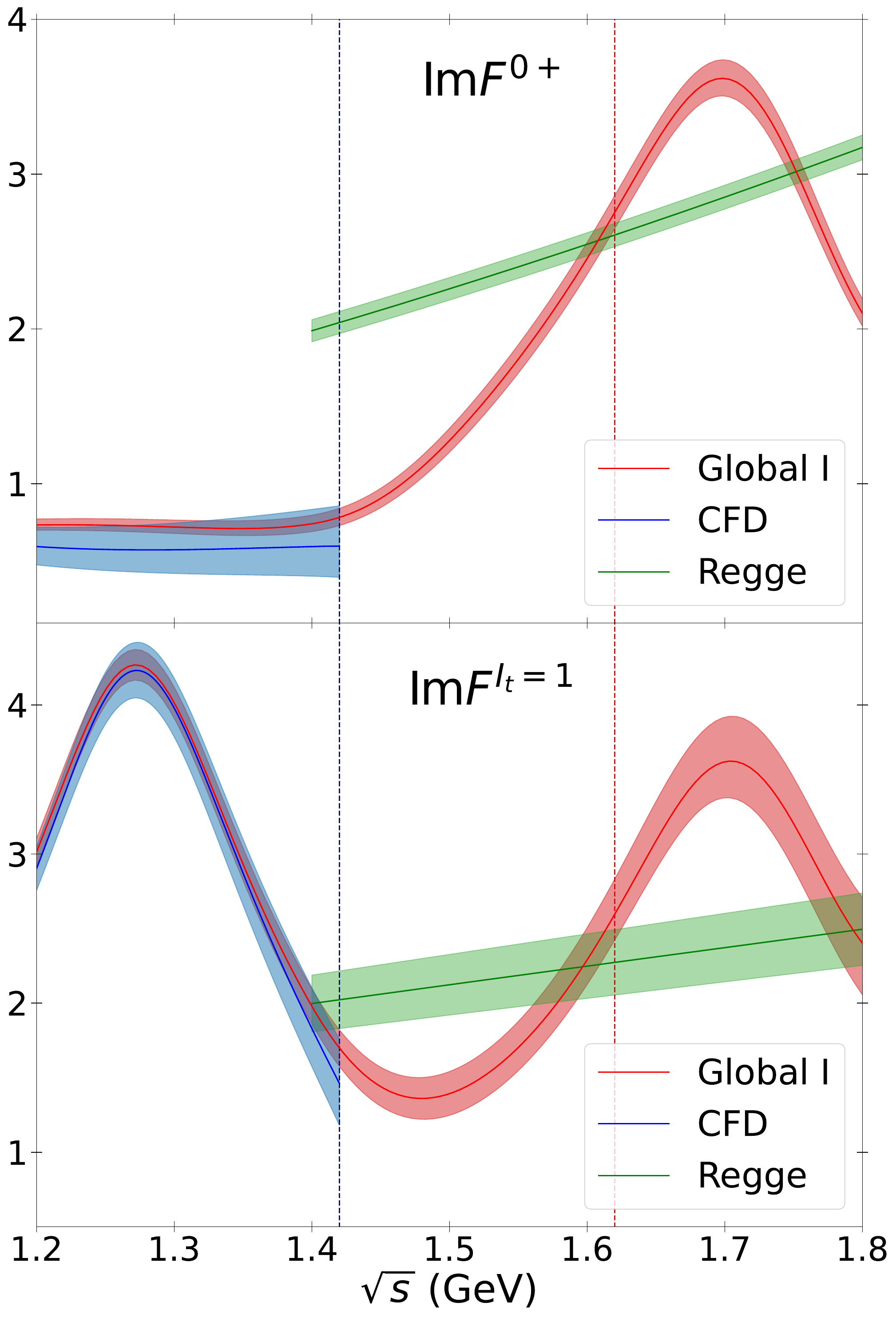}
\caption{ \small \label{fig:improvements} 
Some of the $\pi \pi$ dispersive analysis improvements for precision at high energies \cite{Pelaez:2024uav}: inclusion of the G0-wave (left), new global P-wave (center) and better matching at higher energies with Regge parameterizations (right). These are preliminary results for the so-called Solution I.}
\end{figure*}

However, aiming at precision also above 0.9 GeV, several features of those $\pi \pi$ dispersive amplitude analyses could be improved. First, it is necessary to include the G0 and G2-waves (see Fig. \ref{fig:improvements}) in order to describe the interaction up to higher energies. Second, we must build global parameterizations for S2, D0, D2, F-waves up to 1.8/2 GeV (for isospin $I=2$ data reaches 2 GeV, but for the rest of the partial waves, data do not reach such high energy) improving the data description in certain regions, specially regarding the inelasticities. Furthermore, the P-wave global parameterization has to be improved as well, opening the inelasticity at the $\pi \omega$ instead of the $KK$ threshold (as done in \cite{Pelaez:2019eqa}), and describing the data at higher energies more precisely, as seen in Fig. \ref{fig:improvements}. Finally, the matching of the partial-wave amplitudes with the Regge regime implemented in \cite{GarciaMartin:2011cn} has to be improved and extended to higher energies (Fig. \ref{fig:improvements}) for two of the amplitudes to impose FDR up to higher energies and to avoid the appearance of artifacts.

\begin{figure*}[h]
\centering
\includegraphics[width=0.97\textwidth]{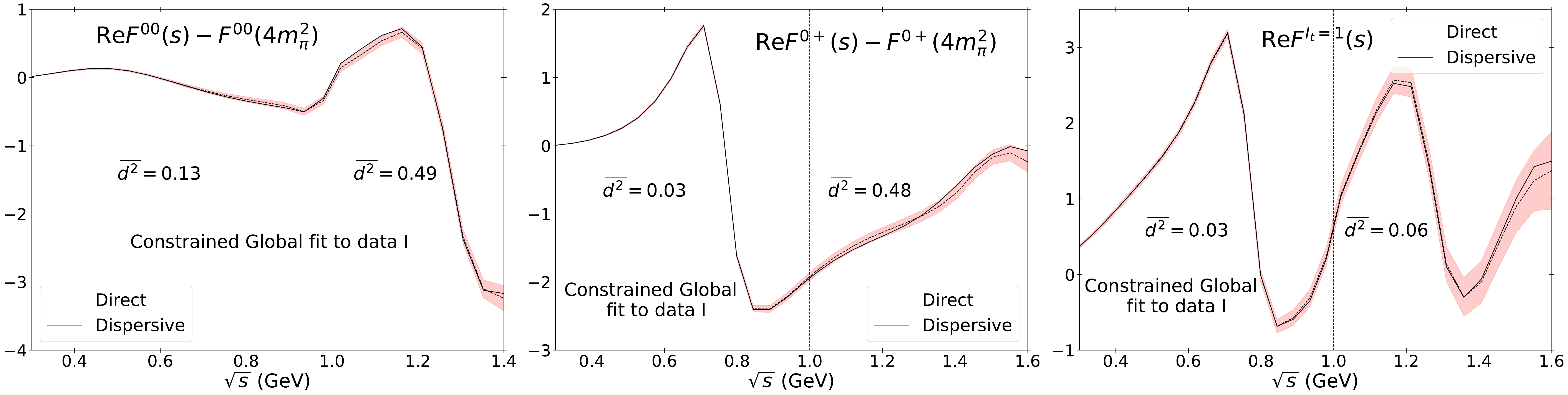}
\includegraphics[width=0.97\textwidth]{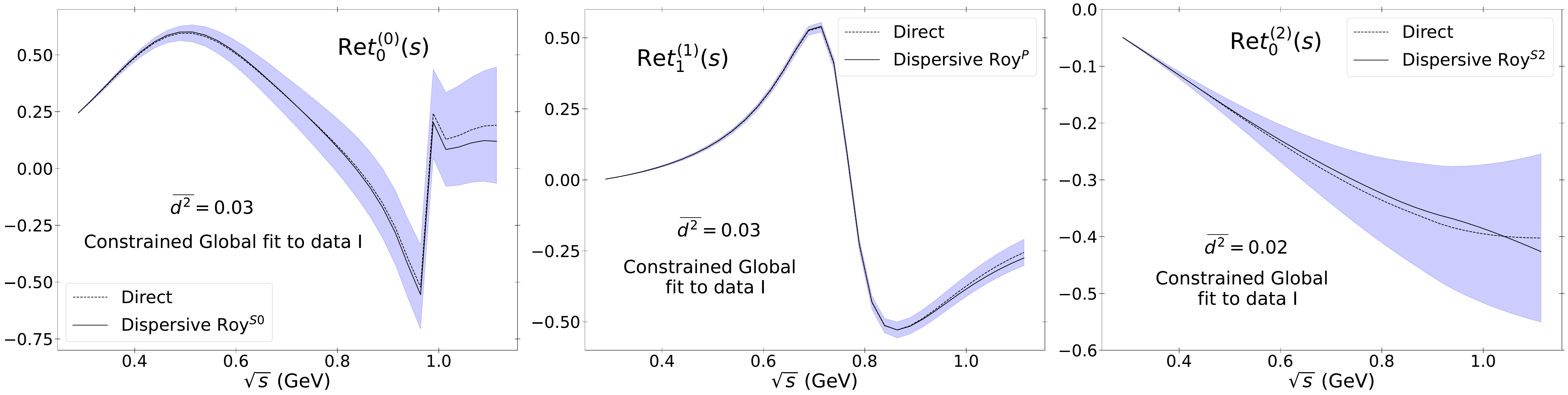}
\includegraphics[width=0.97\textwidth]{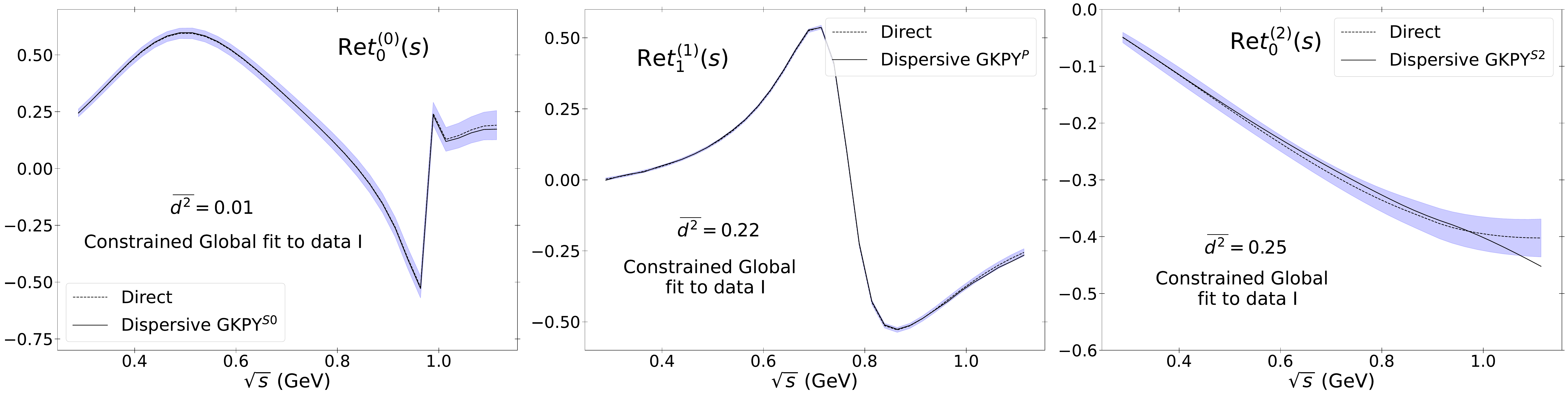}
\caption{ \small \label{fig:disp} 
Preliminary results \cite{Pelaez:2024uav} for the Solution I: fulfillment of the Forward (upper row), Roy (central row) and GKPY (lower row) dispersion relations. The continuous line corresponds to the real part of the dispersion relation obtained from dispersive integrals, the dashed line is calculated directly from the global parameterizations, whereas the band is the error in their difference, attached to the "direct" one for illustration. }
\end{figure*}

Including the new global parameterizations, which describe data up to higher energies, and the improved matching with the Regge regime at higher energies for two of the three amplitudes allow us to impose their FDR up to 1.6 GeV (for one of them, the matching with Regge must remain at 1.4 GeV to avoid artifacts). Additionally, we impose Roy and GKPY dispersion relations up to 1.1 GeV, performing a constrained fit to data with penalty functions of the form
\begin{equation}
    \widebar{d_i^2}=\sum_{k=1}^{N_i} \left(\frac{d_i^k}{\Delta d_i^k}\right)^2,
\end{equation}
where $i$ runs over the three FDR and six Roy-like dispersion relations, $d_i^k$ is the difference of the direct and dispersive real parts at the energies $e_i^k$ on an uniform grid of $N_i$ points. These penalty functions have associated weights so that the final fit has $\widebar{d_i^2}\leq1$ (uniformly in the whole energy regions) for all the dispersion relations. Preliminary results for the Solution I (we study the three most-reliable solutions for the datasets at high energies) can be seen in Fig. \ref{fig:disp}.

\section{Resonance determination}

\begin{figure*}[h]
\centering
\includegraphics[width=0.97\textwidth]{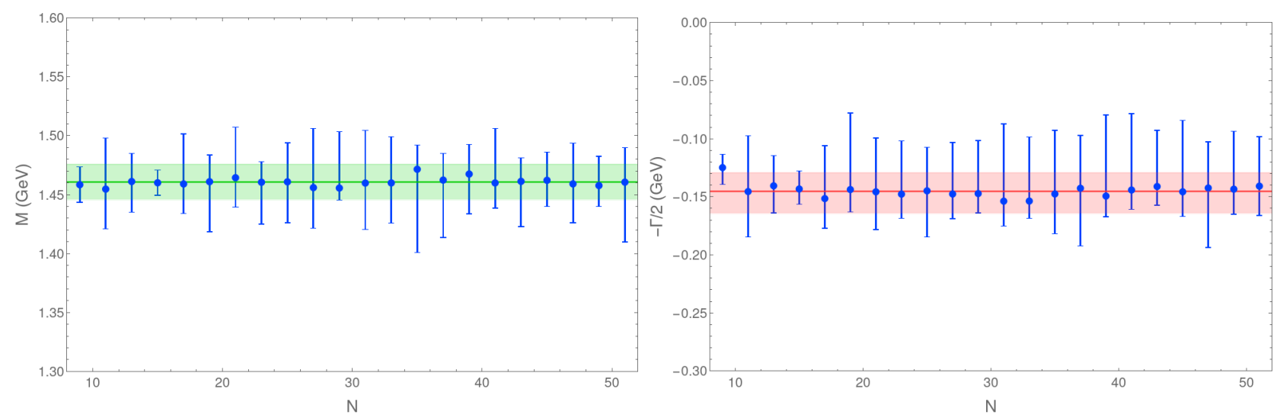}
\caption{ \small \label{fig:1450} 
Preliminary results \cite{Pelaez:2024uav} for the mass (left) and width (right) of the $\rho(1450)$ (Solution I).}
\end{figure*}

With the aim of obtaining pole parameters in a model and parameterization-independent way, we perform an analytic continuation of the output of the FDR from a real segment to the complex plane. For this purpose, we use a robust and general method: continued fractions
\begin{equation}
\label{continued}
    C_{N}(s)=a_0\Big/\Big(1+\frac{a_{1}\left(s-s_{1}\right)}{1+\frac{a_{2}\left(s-s_{2}\right)}{\ddots a_{N-1}\left(s-s_{N-1}\right)}}\Big)\,,
\end{equation}
and the procedure is to interpolate the FDR output calculated at $N$ equally-spaced points in a real segment. Since Eq. \eqref{continued} is a Pad\'e approximant of order $((N-1)/2,(N-1)/2)$ (we choose odd $N$ for technical reasons), it can hold poles and reproduce resonances in the complex plane. For their errors we vary $N$, the real segment and the parameters of our global parameterizations, and perform a weighted mean. The preliminary results for Solution I are stable against $N$, as seen in Fig. \ref{fig:1450} for the $\rho(1450)$ (which we now find due to precision improvements at high energies). Preliminary results for mass, width (defined as $\sqrt{s_{\text{pole}}}=M-i\Gamma/2$) and coupling (as defined in \cite{GarciaMartin:2011jx}) for all the resonances can be found in Table \ref{tab:resonances}. We do not find any hint of the $\rho(1250)$, which was present in old Review of Particle Physics editions and recently claimed \cite{Hammoud:2020aqi} for the same data we fit \cite{Hyams:1973zf,Hyams:1975mc}.

\begin{table}[h]
\renewcommand{\arraystretch}{1.3}
\begin{minipage}{.53\textwidth}
    \resizebox{\textwidth}{!}{
	      \begin{tabular}{|c|c|c|c|} \hline
		\multicolumn{4}{|c|}{\textbf{Isoscalar resonances }}  \\ \hline
		& \textbf{ M} (MeV) & $\boldsymbol{\Gamma}$ (MeV)& $\boldsymbol{|g|}$\\ \hline
		\text{$f_0(500)$} &$457^{+12}_{-10}$ & $518^{+17}_{-26}$& $3.14^{+0.25}_{-0.34}$ GeV \\\hline
		\text{$f_0(980)$} &$990^{+5}_{-6}$ & $44^{+15}_{-10}$& $1.5^{+0.4}_{-0.3}$ GeV\\\hline
		\text{ $f_0(1370)$} &$1211^{+40}_{-42}$ & $582^{+56}_{-85}$& $8.7^{+1.3}_{-1.0}$ GeV \\\hline
		\text{ $f_0(1500)$} &$1547^{+17}_{-15}$ & $55^{+47}_{-29}$ & $5.5^{+0.9}_{-0.8}$ GeV\\\hline
		\text{$f_2(1270)$} &$1265.8^{+0.7}_{-0.5}$ & $196.9^{+0.8}_{-0.7}$ &$4.50^{+0.03}_{-0.05}$ GeV$^{-1}$ \\\hline
    	\end{tabular}}
 \end{minipage}%
\hfill
\begin{minipage}{0.47\textwidth}
    \resizebox{\textwidth}{!}{
	\begin{tabular}{|c|c|c|c|} \hline
		\multicolumn{4}{|c|}{\textbf{Isovector resonances }}  \\ \hline
		& \textbf{ M} (MeV) & $\boldsymbol{\Gamma}$ (MeV)& $\boldsymbol{|g|}$ \\ \hline
		\text{ $\rho(770)$} &$758.0^{+1.1}_{-0.8}$ & $149.0^{+1.0}_{-0.9}$ & $6.062^{+0.012}_{-0.005}$\\\hline
		\text{$\rho(1450)$} &$1459^{+14}_{-11}$ & $278^{+33}_{-36}$ & $1.8{+0.3}_{-0.4}$ \\\hline
		\text{$\rho_3(1690)/$} &  \multirow{ 2}{*}{$1700^{+67}_{-78}$}& \multirow{ 2}{*}{$278^{+174}_{-172}$} & \multirow{ 2}{*}{$XXX$}  \\
  	\text{$\rho(1700)$} & &  &  \\\hline

	\end{tabular}}
\end{minipage}%
\caption{Preliminary pole parameters \cite{Pelaez:2024uav} (mass, width and coupling) for Solution I. The $\rho_3(1690)$ and $\rho(1700)$ overlap in our FDR output and cannot be disentangled (therefore the coupling cannot be calculated).}
\label{tab:resonances}
\end{table}

\section{Summary}

We improve on the $\pi \pi$ dispersive analyses in~\cite{GarciaMartin:2011cn,Pelaez:2019eqa} and obtain a set of global parameterizations that describe data in a more suitable way and satisfy dispersive constraints up to higher energies. Preliminary results show that these constraints are better satisfied for Solution I. From the output of the FDR, we extract the pole parameters using continued fractions, and the results are stable.

\begin{acknowledgments} 
This work is funded by the MCIN/AEI/10.13039/501100011033 (grant PID2022-136510NB-C31) and the European Union’s Horizon 2020 research and innovation program (grant agreement No.824093). P. R. is supported by the Spanish Ministerio de Universidades, fellowship FPU21/03878, and J. R. E. by the Ram\'on y Cajal program (RYC2019-027605-I) of the Spanish MINECO. 
\color{black}
\end{acknowledgments}

\bibliographystyle{JHEP}
\bibliography{largebiblio.bib}

\end{document}